\documentclass[conference]{IEEEtran}
\ifCLASSINFOpdf
   \usepackage[pdftex]{graphicx}
   \usepackage{amsmath}
\else
\fi
\begin{document}
\title{Building Cyber-Physical Energy Systems}

\author{\IEEEauthorblockN{Grigore Stamatescu, Iulia Stamatescu, Nicoleta Arghira, Vasile Calofir, Ioana F\u ag\u ar\u a\c san}
\IEEEauthorblockA{Department of Automatic Control and Industrial Informatics\\
University "Politehnica" of Bucharest, Romania\\
Email: \{grig, iulia, nicoleta, vasile, ioana\}@shiva.pub.ro}
}


%Telephone: (800) 555--1212, Fax: (888) 555--1212}


\maketitle

\begin{abstract}
The built environment, as hallmark of modern society, has become one of the key drivers of energy demand. This makes for meaningful application of novel paradigms, such as cyber-physical systems, with large scale impact for both primary energy consumption reduction as well as (micro-) grid stability problems. In a bottom-up approach we analyze the drivers of CPS design, deployment and adoption in smart buildings. This ranges from low-level embedded and real time system challenges, instrumentation and control issues, up to ICT security layers protecting information in a world of ubiquitous connectivity. A modeling and predictive control framework is also discussed with outlook of deployment for HVAC optimization to a new facility for research from our campus.
\end{abstract}


% \fi
\IEEEpeerreviewmaketitle

\section{Introduction}

By including a vast array of networked measurement, communication and control devices \cite{6931426}, modern buildings offer a fertile ground for blending the physical and virtual worlds under a common paradigm. This is also due to the fact that the various subsystems that enable efficient operation of the building: heating, ventilation and air conditioning (HVAC), power metering, access control \cite{majumdar16} and security, have to offer a unified view to the facilities administrator at the management level. As buildings have been estimated to consume almost 40 \% of the primary energy resources, there is a key incentive for a wide range of stakeholders to contribute to the improvement of system efficiency in buildings. Experimental test-beds can be scaled up and replicated in order to validate the economics of the proposed approaches. 

Cyber-physical systems offer potential for an integrative approach to model, design and implement complex systems at a large scale. Rather than providing some clear design paths, they offer a flexible concept which allows considerable freedom to the designer. Applications of CPS can be already found in the automotive and embedded domains, security, process control and factory automation and others. A critical component has been identified as providing the required security policies and algorithms at the information layer which concern both software and hardware elements. As we are considering large and complex systems, vulnerabilities are able to propagate and cause ireversible damage \cite{paridaricyber}.

Recent references have coined the term cyber-physical energy systems (CPES) \cite{5484019} when discussing the integration of computing, communication and control in the energy domain. This includes generation, both conventional and renewables, transport and distribution, energy storage \cite{6842417}, as well as the consumer side. To test and validate CPES, several test-beds have been implemented which integrate measurement and control hardware, SCADA software frameworks, and a range of communication protocols and networking hardware.
In \cite{6197394} the challenges for sustainable energy grids have been discussed. The key argument revolves around solving the issue of increasing adoption of variable renewable energy sources in connection with energy storage and demand-side strategies. CPES is seen as an essential class of CPS and one that can contribute to this critical challenge through a 3M: monitor, model and mitigate methodology and large-scale control at both ends.

The purpose of this paper is focused on discussing the extension of CPES towards smart buildings. We argue that this approach is viable as one of key elements to drive energy efficiency gains in buildings. Three main pillars related to HVAC control, electrical load monitoring and forecasting and smart grid integration are identified. We propose a case study of how the integration and deployment of a builiding CPES for thermal optimization can be applied in regard to modeling and software toolchain to a new building in our campus.

\section{Building CPES Components}

As fundamental elements of a building CPES paradigm, we identify three main functions where the application of such systems can provide signficant impact from an economic, environmental and social perspective: optimal HVAC control, local management of electric loads and smart grid integration. These are subsequently elaborated upon.

\subsection{HVAC modeling and control}

Control and optimization of the HVAC system is perhaps the most important task of a smart building due to its broad impact on the quality of life of its occupants and to the economic bottom line of the building operator. Small efficiency increments can represent very large cost savings when considering the scale at which these systems operate.

As an important reference, a large scale demonstrator for a real building was put in place in the Opticontrol project \cite{7087366}. The main goal was to design a predictive control strategy for a Swiss office building, to be validated against a more conventional rule-based system already in place.
Alternatively, experimental validation for scenario-based MPC is described in \cite{Parisio2014599}. Cascaded MPC controllers for ventilation requirements (CO2 concentration) and temperature are implemented. These act as high level set-point generators for the lower level PI controllers which are embedded in the air handling units and room radiators. A standard sampling time of 10 minutes is used in conjunction with an 8 hours weather and occupancy prediction horizon. Results show the flexibility of the SMPC strategy in accomodating weather and occupancy uncertainty.

\begin{figure}[ht]
\centering
\includegraphics[width=\columnwidth]{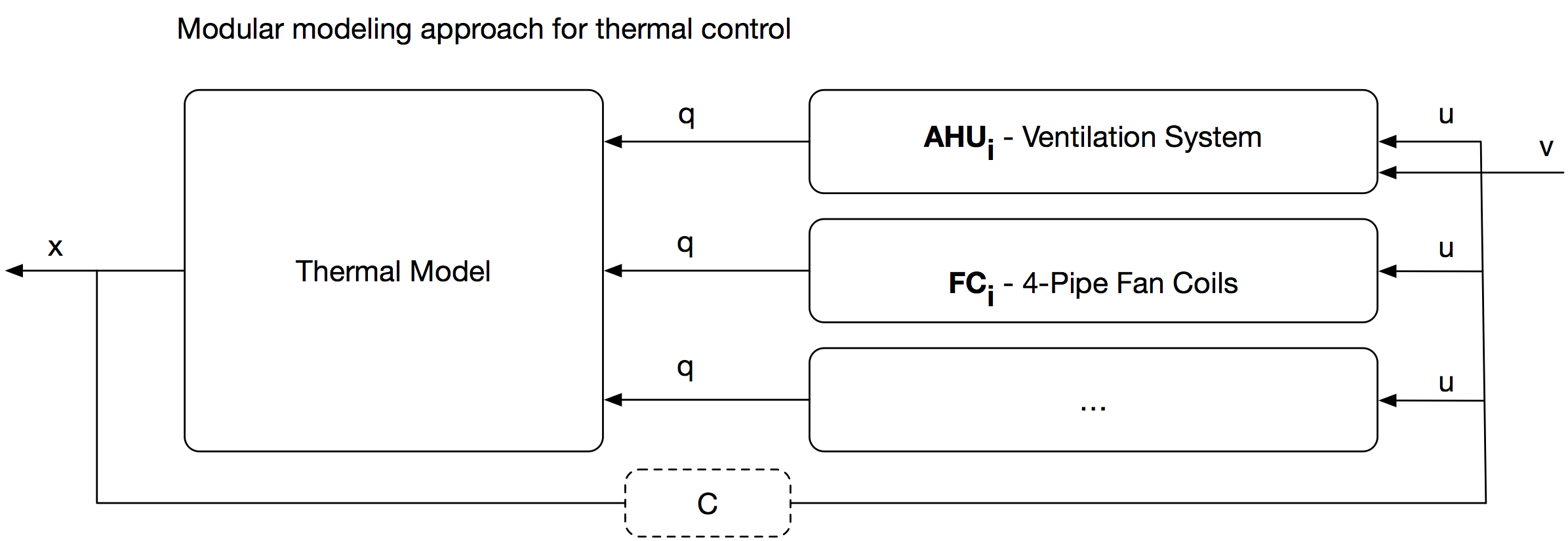}
\caption{Modular modeling aproach for HVAC using BRCM}
\label{bv11}
\end{figure}

An useful tool is the Building Resistive Capacitive Modeling toolbox for MATLAB. This provides a straight forward means to generate simplified thermal models from building information or by directly converting existing models e.g. Energy+ input data files \cite{med16}. The core modeling method by using the BRCM is shown in Figure \ref{bv11}. It relies on the basic idea of separating the core thermal dynamics from the external heat fluxes assigned to the HVAC equipment and system structure, as well as outside and internal influences from weather, occupancy, internal loads, etc.

Based on this high-level concept, we exemplify the state space formulation that is achieved which, in its linearized form, is suitable for predictive controller synthesis.
The full discrete state space model of the building is listed below:

\begin{multline}
x[k+1] = Ax[k] + B_uu[k] + \sum_{i=1}^{n_u}{B_{xu}}_ix[k]u_i[k] + \\ + B_vv[k] + \sum_{i=1}^{n_v}{B_{vu}}_iv[k]u_i[k] + d[k]
\end{multline}

In summary, some key research topics are identified as:

\begin{itemize}
\item simple but accurate thermal models, suitable for real-time optimization;
\item integrated modeling and control framework across the existing infrastructure and toolchain;
\item experimental long-term deployments and validations to highlight cost savings.
\end{itemize}

\subsection{Electrical energy: monitoring and power quality issues}

The monitoring and prediction of electrical energy consumption at the building level has been approached by many researchers. A series of contributions offer software based solutions to deal with improving the energy usage of IT equipment or even whole datacenters. In \cite{5523244} the authors present a method to quantify the energy baseline of a smart building, to which subsequent improvements can be related.

The LoCal project was focused on developing a system framework to aggregate data from multiple heterogeneous measurement points and allow an intelligent decision support system to assint the building manager to optimize energy usage and detect failures.

Power quality issues are investigated at a large scale through complex equipment providing synchronized measurement that are placed at key nodes of the grid. These phasor measurement units (PMUs) are usually installed by utilities and grid operators which are able to justify large investments against grid stability. Recent work \cite{7460660} has approached smaller scale deployments at the microgrid/home level. The system is used to measure and communicate microgrid voltage, frequency and phase angle with reasonable precision and delay over WiFi and LTE links.

Key research topics:

\begin{itemize}
\item energy disaggregation to from power meter data series to pinpoint individual loads and their classification;
\item load forecasting;
\item micro-monitoring of local power quality indicators;
\item new mechansims for user awareness and engagement for energy efficiency.
\end{itemize}

\subsection{Demand response through Smart Grid integration}

At a higher level, considering the fact that a smart building is usually interconnected and carries out bidirectional communication with the electrical grid, algorithms have to be put in place to synchronize it to outside evolutions. Demand response (DR) strategies assume the adjustment of the building's load in reaction to grid conditions e.g. peak periods, where a contractual obligation or economic benefits might bring upon a temporary reduction in the building's energy requirements.

The electrical grid is moving from a load-following to a load-shaping strategy. In this context, methods to control the demand side resources have to be developed. The concept of demand dispatch appears as a complement to the Supply Dispatch. Demand dispatch represents a possible end state that can optimize grid operations beyond what can be achieved with Supply Dispatch alone. Supply Dispatch relies on Ògeneration following the loadÓ while Demand Dispatch allows Òload to follow the generationÓ enabling full optimization of both supply and demand. It enables the management of consumersÕ loads (RES owned by consumers, storage capacities and controllable loads).
The load dispatch is defined as an operating model used by grid operators to dispatch Òbehind-the-meterÓ resources in both directionsÑincreasing and decreasing load as viewed at the system levelÑas a complement to supply (generation) dispatch to more effectively optimize grid operations \cite{Arghira2012128}.

In \cite{BehlJM16} several data-driven methods and tools are discussed for implementation in an open DR assistant tool. Existing basic rule-based approaches are compared to an intelligent method using regression trees to forecast DR events. The types of data that have been used included weather data, scheduling information and building subsystem states.The DIADR project \cite{peffer12} represents a salient proof-of-concept where both open software and hardware components have been integrated into a unitary intelligent energy management system. 

Key research topics:

\begin{itemize}
\item reliable automated demand response algorithms;
\item standardization;
\item building-utility communication.
\end{itemize}

\section{Case study - PRECIS}

\begin{figure}[ht]
\centering
\includegraphics[width=7 cm]{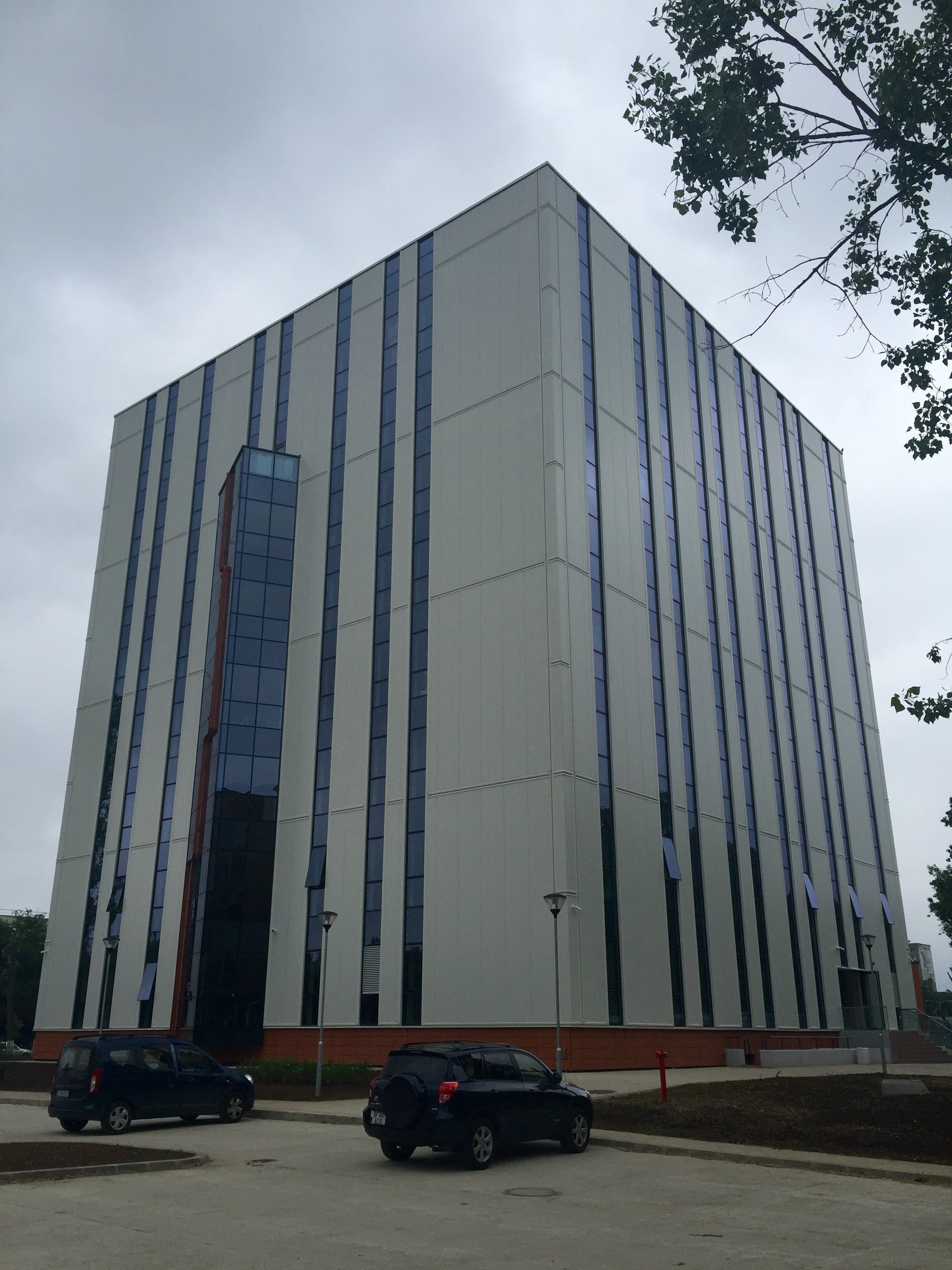}
\caption{Target building - PRECIS research center}
\label{bv1}
\end{figure}

In this section we point out how one the three drivers, namely HVAC control, of building CPES implementations can be applied on a real building. Our target application concerns a new building from our campus, commissioned in December 2015 (Figure \ref{bv1}). It is a 7-floor facility with an overall area of around 9000 sqm which includes research laboratories, office spaces, multifunctional rooms and an auditorium.

\begin{figure}[ht]
\centering
\includegraphics[width=\columnwidth]{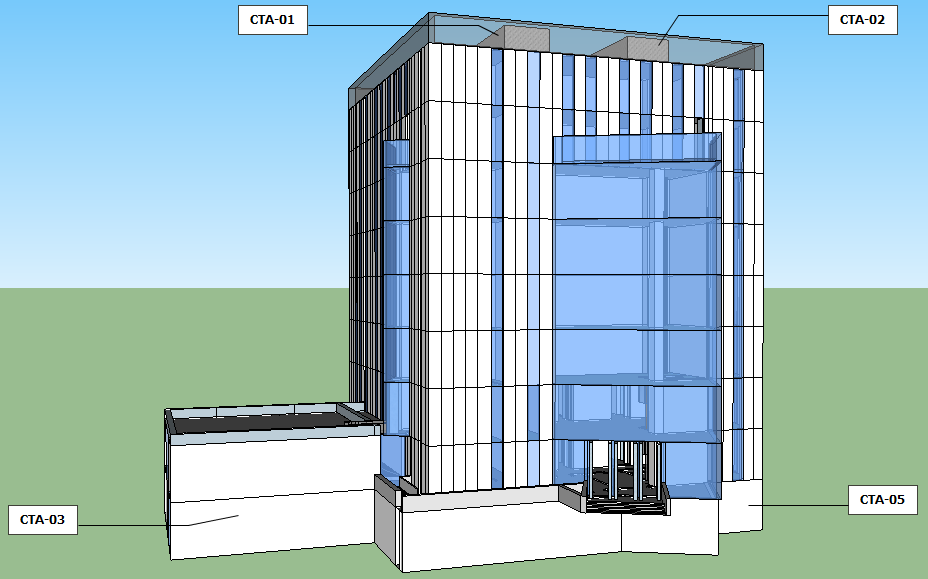}
\caption{Graphical model with AHU placement}
\label{bv1a}
\end{figure}

The HVAC system is composed of a heating/cooling loop and an air loop for ventilation, which work in parallel to achieve the desired comfort set-points. The heating/cooling loop uses chillers for the cold circuit and is connected to the main campus plant at the heating circuit level. Thermal energy is delivered to the zones by means of basic 4-pipe fan coils which rely on three-way valves to mix up the thermal agent and a variable speed fan which blows air over the coil. The air loop satisfies ventilation constraints by bringing in air from the outside, with an economizer function that enables recirculation of the already heated/cooled air for energy efficiency. To a lesser extent, the AHUs are able to also warm up or cool the air if needed.
An existing building management system (BMS) based on Honeywell technology is put into place. It handles the data aggregation from the network of BACNET compatible equipment and controllers, while providing visualization, actuation and alarming functionality in a central location. An important drawback at this point is that control at the room level is local-only with the BMS not being aware of the conditions in the individual thermal zones. Figure \ref{bv1a} highlights the graphical depiction of the building, along with the placement of the five air handling units, CTA1-5.

\begin{figure}[ht]
\centering
\includegraphics[width=\columnwidth]{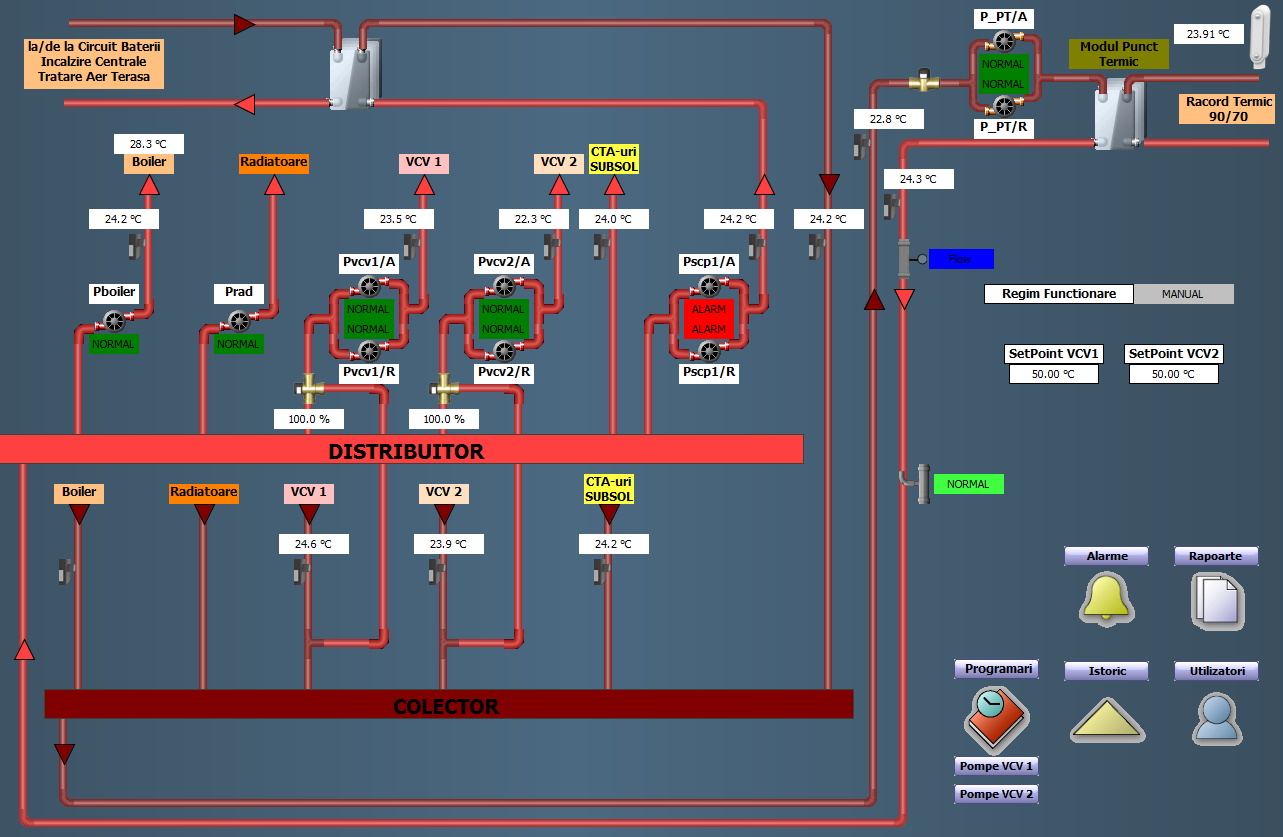}
\caption{Example screen of the current BMS solution - Heating system view}
\label{bv2}
\end{figure}

In Figure \ref{bv2}, the visualization screen for the primary heating loop is presented, starting from the link to the campus-wide heating loop, pumps and measurement points and the interface to the secondary circuits across the floors. Auxiliary systems which use the heating loop are also depicted such as: water heater, several radiators or the air handling units.

\begin{figure}[ht]
\centering
\includegraphics[width=\columnwidth]{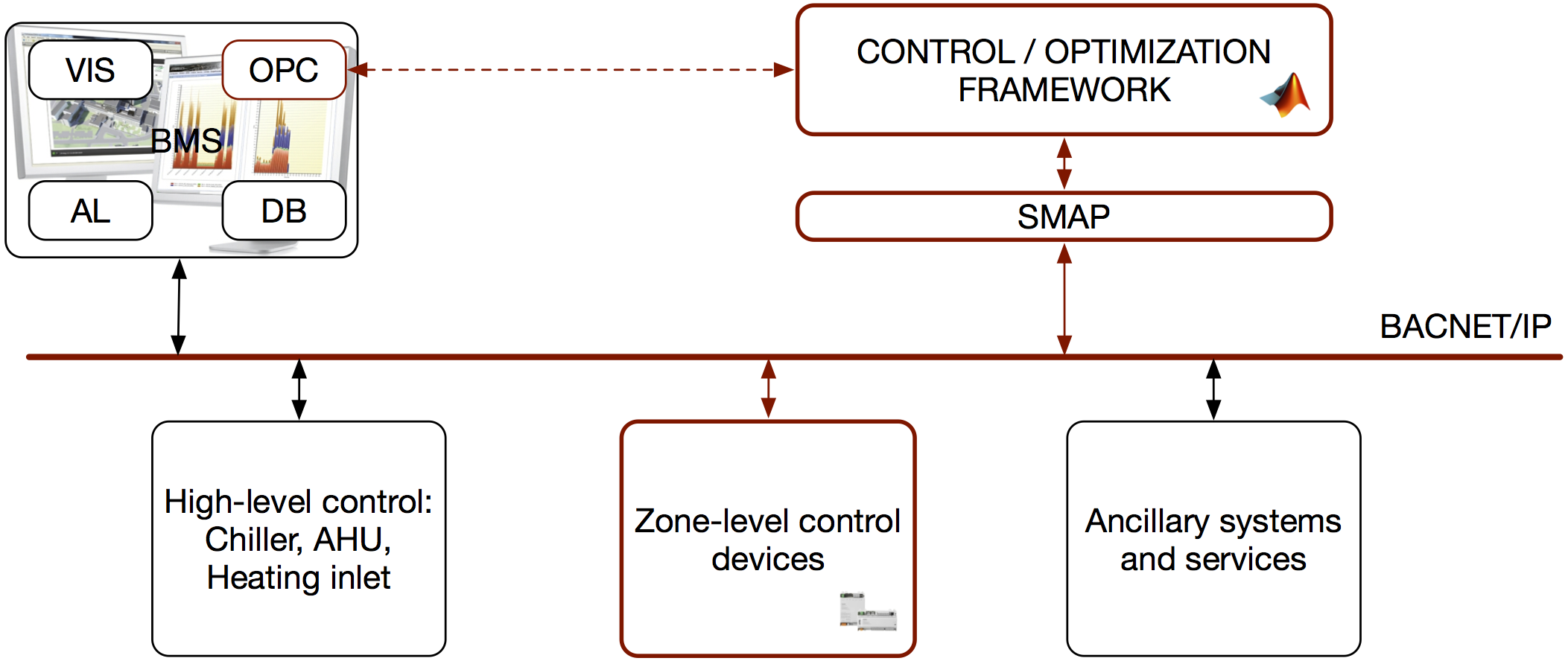}
\caption{System integration via BACNet}
\label{bv3}
\end{figure}

Our proposal for a pilot system implementation for building CPES addresses zone-level control of temperature by designing a supervisor control and optimization framework. The main goal is to formulate the optimization problem for a model predictive controller which minimizes energy cost under a series of constraints: comfort, phyisical, outside conditions. Weather information can be requested from a local network of sensing nodes or using on-line services which can also provide forecast data. As a tool to collect data in real-time from multiple data sources, we use the web-based middleware SMAP \cite{Dawson-Haggerty:2010:SSM:1869983.1870003} which has been proven a reliable, open solution even for very large scale deployments. It provides a library of drivers to poll various instruments like sensors, flow and power meters, and more generic data sources such as OPC servers, but also detailed documentation on writing customized drivers for other equipment. A graphical engine allows visualization of the data streams in real-time. As alternative, data can be exchanged using industry standard protocols with compatible device, through the OPC server at the BMS. The system integration over the communication backbone is shown in Figure \ref{bv3}. The aim is to mirror some of the BMS functionality without interfering with its mission-critical operation. As we initially focus on a subset area for zone-level monitoring and control the proposed system is able to function in parallel and relate to the BMS for energy baselines and validation.

\begin{figure}[ht]
\centering
\includegraphics[width=7cm]{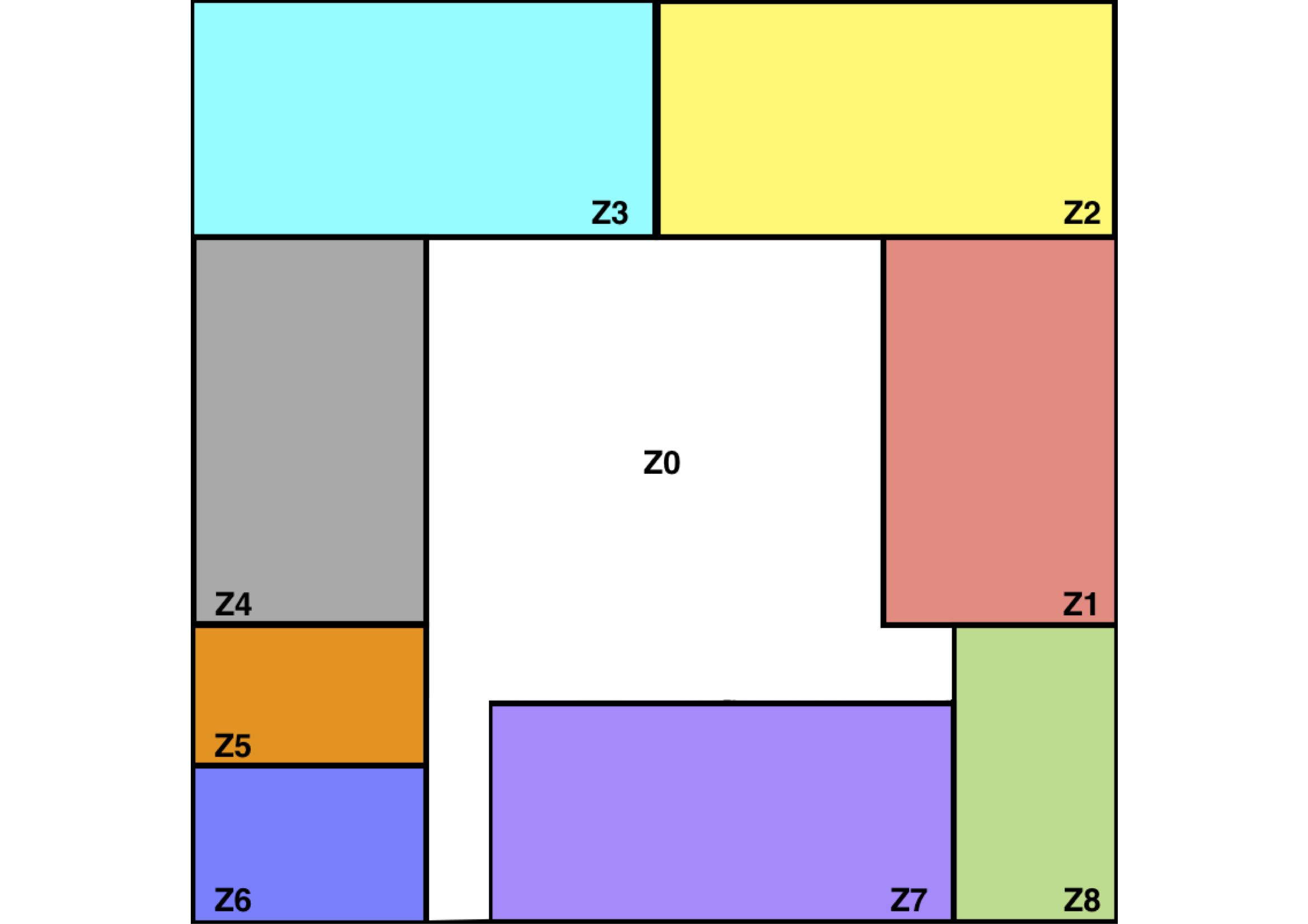}
\caption{Thermal zone layout}
\label{bv4}
\end{figure}

A sample floor layout is provided in Figure \ref{bv4}. The thermal zones i.e. constant temperature volumes of air, are illustrated and correspond to rooms, as follows: zones Z1-Z4 and Z7 are research laboratories, Z5-Z6 are two office spaces, Z8 is a common/meeting room and the zone Z0 accounts for the hallways and auxiliary spaces. The five laboratories are roughly the same area while differing in their orientation which influences the pattern of solar gain throughout the day. By installing room level controllers which measure local parameters: temperature, CO2, window status, and provide the required outputs to actuate the 4-pipe fan coils, local control is implemented. As the controllers are able to communicate over the network, the SMAP server is able to gather data about the room conditions and relay the decision of the control framework to the local actuators. Feedback will be collected from the occupants of the zones in the pilot deployment in order to quantify and improve the acceptance of a centralized strategy for HVAC control.

\section{Conclusion}

The paper discussed the three core elements that enable the CPES paradigm in smart buildings. These concerned thermal and electrical energy management and smart grid integration. As initial step, the proposal for a system architecture has been described to deploy predictive control of temperature at the zone level. Future work is concerned with the design and deployment of a pilot system implementation, accompanied by long-term validation. By federating clusters of smart buildings also a smart campus can be achieved.


\section*{Acknowledgment}

The work has been partially supported by a grant of the Romanian National Authority for Scientific Research, CNDIÐUEFISCDI, project code PN-II-PT-PCCA-2011-3.2-1616.



\bibliographystyle{IEEEtran}
\bibliography{IEEEabrv,bare_conf.bib}

\begin{thebibliography}{10}
\providecommand{\url}[1]{#1}
\csname url@samestyle\endcsname
\providecommand{\newblock}{\relax}
\providecommand{\bibinfo}[2]{#2}
\providecommand{\BIBentrySTDinterwordspacing}{\spaceskip=0pt\relax}
\providecommand{\BIBentryALTinterwordstretchfactor}{4}
\providecommand{\BIBentryALTinterwordspacing}{\spaceskip=\fontdimen2\font plus
\BIBentryALTinterwordstretchfactor\fontdimen3\font minus
  \fontdimen4\font\relax}
\providecommand{\BIBforeignlanguage}[2]{{%
\expandafter\ifx\csname l@#1\endcsname\relax
\typeout{** WARNING: IEEEtran.bst: No hyphenation pattern has been}%
\typeout{** loaded for the language `#1'. Using the pattern for}%
\typeout{** the default language instead.}%
\else
\language=\csname l@#1\endcsname
\fi
#2}}
\providecommand{\BIBdecl}{\relax}
\BIBdecl

\bibitem{6931426}
G.~Stamatescu, C.~Chi{\c t}u, C.~Vasile, I.~Stamatescu, D.~Popescu, and
  V.~Sg{\^a}rciu, ``Analytical and experimental sensor node energy modeling in
  ambient monitoring,'' in \emph{2014 9th IEEE Conference on Industrial
  Electronics and Applications}, June 2014, pp. 1615--1620.

\bibitem{majumdar16}
D.~H.~A. Abhinandan~Majumdar, Zhiru~Zhang, ``Characterizing the benefits and
  limitations of smart building meeting room scheduling,'' in \emph{7th
  International Conference of Cyber-Physical Systems (ICCPS)}, 2016.

\bibitem{paridaricyber}
K.~Paridari, A.~E.-D. Mady, S.~La~Porta, R.~Chabukswar, J.~Blanco, A.~Teixeira,
  H.~Sandberg, and M.~Boubekeur, ``Cyber-physical-security framework for
  building energy management system,'' in \emph{7th International Conference of
  Cyber-Physical Systems (ICCPS)}, 2016.

\bibitem{5484019}
T.~H.~M. et~al., ``Engineering future cyber-physical energy systems:
  Challenges, research needs, and roadmap,'' in \emph{North American Power
  Symposium (NAPS), 2009}, 2009, pp. 1--6.

\bibitem{6842417}
G.~Stamatescu, I.~Stamatescu, N.~Arghira, I.~F{\u a}g{\u a}r{\u a}san, and
  S.~S. Iliescu, ``Embedded networked monitoring and control for renewable
  energy storage systems,'' in \emph{Development and Application Systems (DAS),
  2014 International Conference on}, May 2014, pp. 1--6.

\bibitem{6197394}
J.~Taneja, R.~Katz, and D.~Culler, ``Defining cps challenges in a sustainable
  electricity grid,'' in \emph{Cyber-Physical Systems (ICCPS), 2012 IEEE/ACM
  Third International Conference on}, April 2012, pp. 119--128.

\bibitem{7087366}
D.~Sturzenegger, D.~Gyalistras, M.~Morari, and R.~S. Smith, ``Model predictive
  climate control of a swiss office building: Implementation, results, and cost
  -benefit analysis,'' \emph{IEEE Transactions on Control Systems Technology},
  vol.~24, no.~1, pp. 1--12, Jan 2016.

\bibitem{Parisio2014599}
A.~Parisio, D.~Varagnolo, M.~Molinari, G.~Pattarello, L.~Fabietti, and K.~H.
  Johansson, ``Implementation of a scenario-based mpc for hvac systems: an
  experimental case study,'' \emph{IFAC Proceedings Volumes}, vol.~47, no.~3,
  pp. 599 -- 605, 2014.

\bibitem{med16}
G.~Stamatescu and I.~Stamatescu, ``Open and closed loop simulation for
  predictive control of buildings,'' in \emph{24th Mediterranean Conference on
  Control and Automation (MED'16)}, 2016.

\bibitem{5523244}
J.~Kleissl and Y.~Agarwal, ``Cyber-physical energy systems: Focus on smart
  buildings,'' in \emph{Design Automation Conference (DAC), 2010 47th
  ACM/IEEE}, June 2010, pp. 749--754.

\bibitem{7460660}
Z.~Huang, T.~Zhu, H.~Lu, and W.~Gao, ``Accurate power quality monitoring in
  microgrids,'' in \emph{2016 15th ACM/IEEE International Conference on
  Information Processing in Sensor Networks (IPSN)}, 2016.

\bibitem{Arghira2012128}
N.~Arghira, L.~Hawarah, S.~Ploix, and M.~Jacomino, ``Prediction of appliances
  energy use in smart homes,'' \emph{Energy}, vol.~48, no.~1, pp. 128 -- 134,
  2012.

\bibitem{BehlJM16}
\BIBentryALTinterwordspacing
M.~Behl, A.~Jain, and R.~Mangharam, ``Data-driven modeling, control and tools
  for cyber-physical energy systems,'' \emph{CoRR}, vol. abs/1601.05164, 2016.
  [Online]. Available: \url{http://arxiv.org/abs/1601.05164}
\BIBentrySTDinterwordspacing

\bibitem{peffer12}
T.~P. et~al., ``Deep demand response: The case study of the citris building at
  the university of california-berkeley,'' in \emph{ACEEE Summer Study on
  Energy Efficiency in Buildings}, 2012.

\bibitem{Dawson-Haggerty:2010:SSM:1869983.1870003}
S.~Dawson-Haggerty, X.~Jiang, G.~Tolle, J.~Ortiz, and D.~Culler, ``smap: A
  simple measurement and actuation profile for physical information,'' in
  \emph{Proceedings of the 8th ACM Conference on Embedded Networked Sensor
  Systems}, 2010, pp. 197--210.

\end{thebibliography}

\end{document}